\documentclass[aps,pra,twocolumn]{revtex4-2}
\usepackage[dvipdfmx]{graphicx}				
\usepackage{amssymb,amsmath}
\usepackage{physics}
\usepackage{subfigure}
\usepackage{hyperref }
\usepackage{ulem}
\usepackage{color}
\usepackage{comment}

\definecolor{mycolor1}{rgb}{ 0, 0.8, 0.6}

\begin{document}
\title{Polarizing electron spins with a superconducting  flux qubit }
\author{Shingo Kukita$^{1)}$}
\email{toranojoh@phys.kindai.ac.jp}
\author{Hideaki Ookane$^{1)}$}
\email{h-okane@phys.kindai.ac.jp}
\author{Yuichiro Matsuzaki$^{2)}$}
\email{matsuzaki.yuichiro@aist.go.jp}
\author{Yasushi Kondo$^{1)}$}
\email{ykondo@kindai.ac.jp}
\affiliation{$^{1)}$Department of Physics, 
Kindai University, Higashi-Osaka 577-8502, Japan}
\affiliation{$^{2)}$Device Technology Research Institute,
National Institute of Advanced Industrial Science and Technology (AIST),
1-1-1, Umezono, Tsukuba, Ibaraki 305-8568, Japan}

\begin{abstract}
Electron spin resonance (ESR) is a useful tool  to investigate properties of 
materials in magnetic fields 
where high spin polarization of target electron spins is required in order 
to obtain high sensitivity. 
However, the smaller magnetic fields  becomes, the more difficult high polarization 
is passively obtained by thermalization. 
Here, we propose to employ a superconducting flux qubit (FQ) to polarize electron spins actively. 
We have to overcome a large energy difference between the FQ and electron spins
for  efficient energy transfer among them. 
For this purpose, we adopt a spin-lock technique on the FQ where the Rabi frequency 
associated with the spin-locking can match the resonance (Larmor) one of the electron spins. 
We find that adding dephasing on the spins is beneficial to obtain high polarization of them, 
because otherwise the electron spins are trapped in dark states 
that cannot be coupled with the FQ.
We show that our scheme can achieve high polarization of electron spins 
in realistic experimental conditions. 
\end{abstract}
\maketitle

\section{Introduction}
Increasing attention has been paid to electron spin resonance (ESR) due to an excellent sensitivity
compared with that of nuclear magnetic resonance (NMR).
An improvement of the ESR sensitivity  is important for practical applications.
Therefore,
superconducting circuits have often been used to detect the small number of electron spins \cite{kubo2012electron,toida2016electron,bienfait2016reaching,eichler2017electron,probst2017inductive,bienfait2017magnetic,budoyo2018electron,toida2019electron}.
By using a superconducting resonator, it is possible to measure only $12$ spins 
with 1~s measurement time  
where the detection volume is around 6fl \cite{ranjan2020electron} where 
the frequency of the superconducting resonator is fixed. 
It is favorable to sweep not only the microwave frequency but also the magnetic field
to investigate complicated spin systems. 
For this purpose, we could use a waveguide
\cite{wiemann2015observing}, a frequency tunable resonator \cite{chen2018hyperfine}, or a
direct current-superconducting quantum interference device (dcSQUID) \cite{toida2016electron,yue2017sensitive}.
Among these approaches, a superconducting flux qubit (FQ) is promising  and 
has already achieved a sensitivity of $20$ spins/Hz$^{1/2}$ 
with a sensing volume of 6 fl for the ESR \cite{budoyo2020electron}.

It is worth mentioning that the FQ cannot work if we apply high magnetic fields, 
and so the applied field should be smaller than 10 mT \cite{toida2016electron}.
One of the problems in FQ-ESR  measurements is a low polarization of target electrons especially 
when they are in a low magnetic field. 
A typical thermal energy 
$\sim k_{\rm B} T$ ($k_{\rm B}$: the Boltzmann constant and T: temperature)
at mK temperatures is around hundreds of MHz in frequency unit, while the typical magnetic 
energy of the electron spins $\sim \mu_{\rm B}B$ ($\mu_{\rm B}$: the Bohr magneton constant 
and $B$: flux density in T unit) in a small field of few mT is about tens of MHz. 
This implies that the electron spins cannot be fully polarized in these conditions
and that the sensitivity of ESR is deteriorated.
 Note that high spin polarization of target electrons is required in order to obtain high sensitivity \cite{wertz2012electron}.
A Purcell effect \cite{purcell1995spontaneous} was recently employed 
to polarize electron spins with a superconducting cavity 
\cite{bienfait2016controlling}. However, this is not applicable to the case 
when electron spins are placed in a low magnetic field because of a large energy difference between the cavity and electron spins. 
Moreover, a thermal relaxation time of electron spins becomes larger at lower temperature, and 
thus it is difficult to polarize them 
\cite{T1AmsussCavity2011,T1ProbstRareEarth2013,T1AngereBIstable2017,T1BudoyoBottleSCI2018}.

Here, we propose to employ a FQ for not only detecting but also polarizing electron spins.
The main idea is that the energy relaxation time of the FQ is 
much shorter than that of the electron spins, and so we can efficiently emit 
the energy of the electron spin to the environment
by using a coupling between the FQ and electron spins.
We adopt a spin-lock technique where the Rabi frequency of the FQ in a rotating frame 
associated with the spin-locking matches with the resonance (Larmor) frequency of the electron spins 
in a low magnetic field \cite{hartmann1962nuclear}. 
The important difference from the polarization with a Purcell effect \cite{bienfait2016controlling}
is that the Rabi frequency can be much smaller than the resonance one of the FQ. 
By using a long-lived FQ such as a capacitively shunted FQ 
whose
coherence time is 
around tens of micro seconds \cite{you2007low,yan2016flux,abdurakhimov2019long}, 
the Rabi frequency can be reduced to hundreds of kHz.
With these properties, one may overcome the energy scale mismatch between a FQ and 
electron spins, and thus  the efficient polarization of the electron spins becomes possible.   

This paper is organized as follows. 
\S~\ref{sec:theory} illustrates our setup and proposal 
with analytical discussion on a simplified model. We show our numerical simulations 
with realistic experimental parameters in \S~\ref{sec:NS} for demonstrating 
its experiment feasibility. We conclude this paper in \S~\ref{summary}.

\section{Theory}
\label{sec:theory}
We here propose to employ a Hartmann-Hahn (H-H) resonance \cite{hartmann1962nuclear} 
to polarize electron spins with a FQ.  The H-H resonance has been applied to polarize 
environmental
spins by using  nitrogen vacancy (NV) centers in diamond \cite{laraoui2013approach,scheuer2016optically}. Our proposal 
is expected
to polarize far more electron spins than the case of the NV center
in diamond, 
because the size of the FQ is of the order of micrometers while that of the NV center is of the order of angstroms.

We discuss a simplified model in order to illustrate our proposal 
after introducing a Hamiltonian and Lindbladian that govern the electron spins and FQ.

\subsection{Model}
\label{subsec:hamiltonian}
The Hamiltonian of a FQ  coupled to $M$ electron spins (labeled with $k=1 \sim M$) 
is described as follows.
\begin{eqnarray}
	H&=&H_{\rm{FQ}}+H_{\rm{spin}}+H_{\rm{I}}, \nonumber 
\end{eqnarray}
where $H_{\rm{FQ}}$, $H_{\rm{spin}}$, and $H_{\rm{I}}$ denote the Hamiltonian of the FQ, spins, and interaction between them. $H_{\rm{FQ}}$ is given as
\begin{eqnarray*}
	H_{\rm{FQ}}&=&\frac{\epsilon}{2}Z + \frac{\Delta}{2}X  + \lambda Y  \cos \omega t,
\end{eqnarray*}
where $\epsilon$ denotes the energy bias, 
$\Delta $ 
the tunneling energy, 
$\omega$ 
the frequency of the microwave,  
and $\lambda $
the strength of the microwave.
$X$, $Y$, and $Z$
are standard Pauli matrices acting on the FQ.
It is convenient for us to change the notation, and we rewrite $H_{\rm{FQ}}$ as follows.
\begin{eqnarray*}
	H_{\rm{FQ}}&=&\frac{\epsilon}{2}\sigma^{(0)}_x + \frac{\Delta}{2}\sigma^{(0)}_y  
	+ \lambda \sigma_z ^{(0)} \cos \omega t,
\end{eqnarray*}
where we change $X$, $Y$, and $Z$ to $\sigma^{(0)}_{y}$, $\sigma^{(0)}_{z}$ and 
$\sigma^{(0)}_{x}$, respectively. 0 denotes the 0th qubit in our system. 
$H_{\rm{spin}}$ and $H_{\rm{I}}$ are given as 
\begin{eqnarray*}
	H_{\rm{spin}}&=& \sum _{k=1}^M \omega _k \sigma _z ^{(k)}, \\
	H_{\rm{I}}&=& \sum _{k=1}^M g_k \sigma^{(0)}_x \sigma _x ^{(k)} ,
\end{eqnarray*}
where $\omega _k$  denotes the resonance frequency of the $k$-th spin 
and $g_k$ the coupling strength between the FQ and the $k$-th spin. 
By going to a rotating frame with a frequency of $\omega = \sqrt{\epsilon^2+\Delta ^2}$ of the FQ, 
we obtain the following Hamiltonian with a rotating wave approximation
with a condition of $\epsilon \gg \Delta$
\begin{align}
	H\simeq \frac{\lambda}{2} \sigma^{(0)}_z +  \sum _{k=1}^M \omega _k \sigma _z ^{(k)}
	+ \sum _{k=1}^M g_k \sigma^{(0)}_x \sigma _x ^{(k)}.
\end{align}

We obtain the following effective Hamiltonian in a rotating frame of which frequency is $\displaystyle \omega_{\rm avg}= \frac{1}{M} \sum_k \omega _k$ with the rotating wave approximation.
\begin{align}
	\label{eq:ham_FQ_e}
	H&\simeq 
	\sum_{k=1}^M  \left( \omega _k' \sigma _z^{(k)}+
	g_k\left( \sigma^{(0)}_{+} \sigma_{-}^{(k)}
	+\sigma^{(0)}_{-}\sigma_{+}^{(k)} \right) \right),   
\end{align}
where $\omega _k' = \omega_k - \omega_{\rm avg}$ and 
$\sigma_\pm = \sigma_x \pm i \sigma_y$. 
Here, we set $\lambda/2$ to be $\omega_{\rm avg}$.

The energy exchanges occur between the FQ and electron spins 
during the irradiation of a microwave due to the flip-flop interaction, 
while there is no coupling between them in the absence of the irradiation due to the energy detuning of 
\mbox{$\sqrt{\epsilon^2 + \Delta ^2} \gg \omega_k$}.


We also introduce 
the Lindblad operator
in order to describe the
relaxation of 
the system (= a  FQ and spins), as follows. 
\begin{align}
	\label{eq:lin_FQ_e}
	{\cal L}[\rho] 
	=&\sum_{l=0}^M   \gamma_{\rm T}^{(l)} \left(\sigma^{(l)}_{z}\rho \sigma^{(l)}_{z}- \rho \right)
	\nonumber \\
	&+ \sum_{l=0}^M\gamma_{\rm L}^{(l)}\left( \sigma^{(l)}_{+}\rho \sigma^{(l)}_{-} +\sigma^{(l)}_{-}\rho \sigma^{(l)}_{+} - \rho \right),
\end{align}
where $\gamma_{\rm T}$ and $\gamma_{\rm L}$ characterize the strengths of 
transversal and longitudinal relaxations, respectively.
Note that the superscript $l$ runs from $0$ to $M$ while $k$ runs from $1$ to $M$.
We consider the case when
each
qubit has a different relaxation parameter labeled with $*^{(l)}$.
  
A system dynamics is then determined by 
\begin{equation}
	\frac{d \rho}{d t}=-i [H,\rho]+{\cal L}[\rho],
	\label{eq:dynamics}
\end{equation}
while the initial state is assumed to be 
\begin{equation}
	\rho(0)=|0\rangle\langle0|\otimes \bigotimes^{M}_{k=1}\frac{\sigma_{0}}{2},
	\label{eq:realini}
\end{equation}
where $\sigma_{0}$ is the $2 \times 2$ identity matrix. Then, our goal is  to obtain 
\begin{equation}
	\rho({\rm final})=|0\rangle\langle0|\otimes \bigotimes^{M}_{k=1}|0\rangle\langle0|,
\end{equation}
after some operations. 

\subsection{Simplified Model}

We consider a simplified model where $\omega _k$'s and $g_{k}$'s are identical for $k=1 \sim M$. 
Due to this simplification, we can calculate  polarization dynamics for 
a large number of spins.
Our procedure consists of two steps, Step~I and II. 
A FQ interacts with the spins and absorbs their entropy in Step~I
while the spin states are homogenized with a help of dephasing in Step~II. 

\begin{itemize}
	\item in Step~I
	
	Let the system develop according to the following simplified Hamiltonian. 
	This is obtained from Eq.~\eqref{eq:ham_FQ_e} by assuming $g_k = g$ and $\omega_k' = 0$ 
	for $k =1 \sim M$. 
	\begin{align}
		\label{eq:ham_I}
		H^{\rm I}&=g\left( \sigma^{(0)}_{+} S_{-}+\sigma^{(0)}_{-}S_{+} \right), \\
		S_{x, y,z,\pm} &=\sum^{M}_{k=1}\sigma^{(k)}_{x,y,z,\pm}, \nonumber 
	\end{align}
   while the Lindbladian~\eqref{eq:lin_FQ_e} is simplified as
   \begin{equation}
   	\label{eq:lin_I}
   	{\cal L}^{\rm I}[\rho]=\gamma \left(\sigma^{(0)}_{z}\rho \sigma^{(0)}_{z}-\rho \right),
   \end{equation}
	where we assume that only $\gamma_{\rm T}^{(0)} = \gamma \ne 0$ and the other 
	$\gamma_{\rm T}^{(k)}$ and $\gamma_{\rm L}^{(l)}$ are negligible. 
	The only 0th qubit is  under influence of dephasing. 
	
	After the dynamics, we initialize the 0th
	qubit to the ground state $|0\rangle $
	without disturbing the others ($k =1 \sim M$). 

	\item in Step~II
	
	We decouple the 
	0th qubit from the others. 
	Therefore, the Hamiltonian is given as, 
	\begin{align}
	\label{eq:ham_II}
	H^{\rm II} &=0,  
	\end{align}	
	while the Lindbladian~\eqref{eq:lin_FQ_e} in Step~II is simplified as
	\begin{equation}
		 \label{eq:lin_II}
		{\cal L}^{\rm II}[\rho]
		=\sum^{M}_{k=1}\gamma' \left(\sigma^{(k)}_{z}\rho\sigma^{(k)}_{z}-\rho \right), 
	\end{equation}
	where we assume that only $\gamma_{\rm T}^{(k)} = \gamma' \ne 0$ and the other 
	$\gamma_{\rm T}^{(0)}$ and $\gamma_{\rm L}^{(l)}$ are 
	negligible.
	All qubits except the 0th one
	are under influence of the same dephasing. 
	
\end{itemize}

\subsection{Step~I}
We employ the Young-Yamanouchi basis $|j,m,i\rangle$ \cite{ping2002group} in order to represent
the spin state.  Note that 
$j=1/2,3/2,\cdots,M/2$ and $|m|\leq j$ (half-integer) for odd $M$ cases 
while $j=0,1,\cdots,M/2$ and $|m|\leq j$ (integer) for the even $M$ cases.
The index $i$ represents the number of ways of composing $n$ qubits to obtain the total angular momentum $j$ and takes $1\sim d_{j}$, where $d_{j}:=(2 j + 1)M!/(M/2 + j + 1)!(M/2 - j)!$.
The action of spin operators $S_{\pm,z}$ is given as follows. 
\begin{align}
	S_{+}|j,m,i\rangle&=\sqrt{j(j+1)-m(m+1)} |j,m+1,i\rangle, \nonumber\\
	S_{-}|j,m,i\rangle&=\sqrt{j(j+1)-m(m-1)} |j,m-1,i\rangle, \nonumber\\
	\frac{S_{z}}{2}|j,m,i\rangle&=m |j,m,i\rangle.
\end{align}
Let us define 
\begin{align}
	\label{eq:spin_act}
	|a_{jmi} \rangle &:= |0\rangle \otimes |j,m,i \rangle, \nonumber \\
	|b_{jmi} \rangle &:= |1\rangle \otimes |j,m-1,i \rangle.
\end{align}
By using the above bases, the initial state~\eqref{eq:realini} can be rewritten as
\begin{align}
	\rho(0) 
	 &=\frac{1}{2^{M}}\sum_{j,m,i}|a_{j m i}\rangle\langle a_{j m i}|.
\end{align}
The dynamics of $\rho(t)$ from the above initial state according to Eq.~\eqref{eq:dynamics} 
with Eqs.~\eqref{eq:ham_I} and \eqref{eq:lin_I} is easily obtained with the help of Eq.~\eqref{eq:spin_act}:
\begin{align}
    \rho(t)&=\frac{1}{2^{M}}\sum_{j,m,i} \rho_{jmi}(t),
    \nonumber \\
    \rho_{jmi}(t) &:= a_{jmi}(t) |a_{jmi} \rangle \langle a_{jmi}| + b_{jmi}(t) |b_{jmi} \rangle \langle b_{jmi} |
    \nonumber \\
    &+c_{jmi}(t) |a_{jmi} \rangle \langle b_{jmi}| + c_{jmi}^*(t) |b_{jmi} \rangle \langle a_{jmi} |.
\end{align}
Here the coefficients $a_{j m i}(t)$, $b_{j m i}(t)$, $c_{j m i}(t)$ satisfy the following differential equations,
\begin{align}
	\dot{a}_{j m i}(t)=&-2g l_{j m}c_{j m i}^{I}(t),\nonumber\\
	\dot{b}_{j m i}(t)=&2g l_{j m}c_{j m i}^{I}(t),\nonumber\\
	\dot{c}_{j m i}^{I}(t)=&g l_{j m} a_{j m i}(t)-g l_{j m} b_{j m i}(t)-2 \gamma c_{j m i}^{I}(t),\nonumber\\
	\dot{c}_{j m i}^{R}(t)=&-2 \gamma c_{j m i}^{R}(t),
\end{align}
where $c^{I~(R)}_{j m i}(t)$ is the imaginary (real) part of $c_{j m i}$ and $l_{j m}=\sqrt{j(j+1)-m(m-1)}$.
The dynamics of each $\rho_{jmi}(t)$, or the dynamics of $\{a_{j m i}(t),b_{j m i}(t), c_{j m i}(t) \}$, is independent of each other.

Let us now focus on the dynamics of $\rho_{jmi}(t)$ for fixed $j,m,i$.
Because the dynamics of $c_{j m i}^{R}(t)$ is decoupled from those of the other variables, 
we assume that $c_{j m i}$ is pure imaginary from now on. 
The eigenvalues of this dynamics (= decay rates) are given as 
\begin{equation}
	0,~\frac{ -\gamma + \sqrt{\gamma^{2}-(16 g l_{j m})^{2}}}{2},
	~\frac{ -\gamma - \sqrt{\gamma^{2}-(16 g l_{j m})^{2}}}{2}.
\end{equation}
The eigenstate corresponding to the eigenvalue 0 is 
$\frac{1}{2} |a_{j m i}\rangle\langle a_{j m i}|+\frac{1}{2} |b_{j m i}\rangle\langle b_{j m i}|$
and is independent of $\gamma$.
Note that this state corresponds to the fully mixed state in the space spanned 
by $|a_{j m i}\rangle$ and $|b_{j m i}\rangle$, and is stationary. 


Let us consider the dynamics of which initial state is given as 
\begin{equation}
	\rho=\sum_{j,m,i}p_{j,m} |a_{j m i}\rangle\langle a_{j m i}|.
	\label{eq:ini}
\end{equation}
The states at the beginning and the end of Step I + II (and also the initial state) 
can be always written in the above form as shown below. 
The reason why $p_{j,m}$ is independent of the index $i$ is that there is no way to control 
the freedom of $i$ in our protocol and the initial state is also set to be independent of the index $i$.
Because the dynamics of each $|a_{j m i}\rangle\langle a_{j m i}|$ is independent of each other, 
the dynamics from the initial state~(\ref{eq:ini}) is simply given as
\begin{equation}
    \rho(t)=\sum_{j,m,i} p_{j,m}\rho_{jmi}(t).
\end{equation}
Then, it is assumed that we can wait until the above dynamics converges.
After that, we obtain the stationary state 
of which  density matrix $\rho_{\rm st}$ is given as 
\begin{align}
	\rho_{\rm st} 
	&=\sum_{j,m,i}\frac{p_{j,m}}{2}(|a_{j m i}\rangle 
	\langle a_{j m i}|+|b_{j m i}\rangle \langle b_{j m i}|).
\end{align}

\begin{figure}[t]
	\begin{center}
		\includegraphics[width=9.5cm]{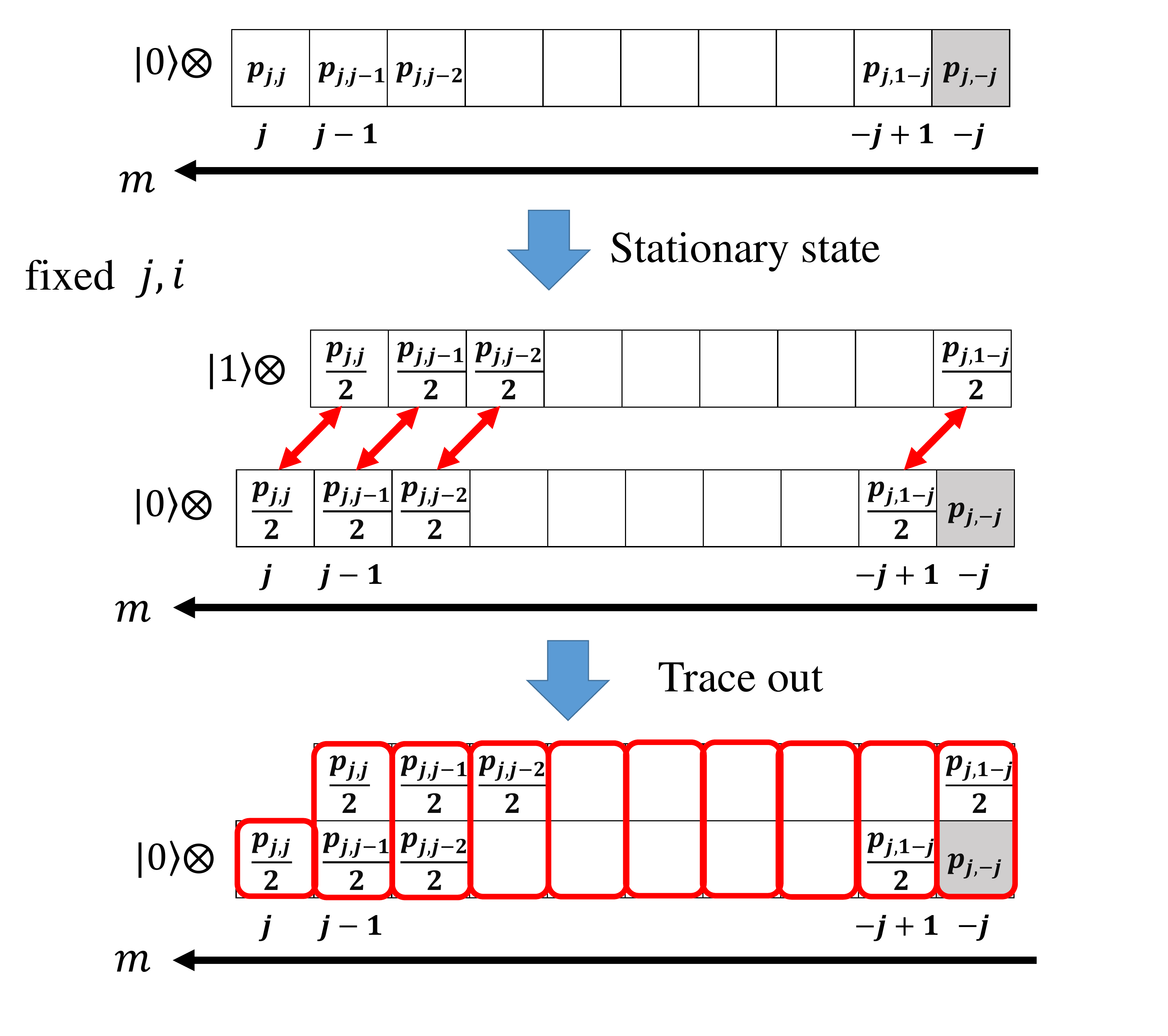}
	\end{center}
	\caption{(color on line) Schematic picture of Dynamics of Step~I which 
		consists of ``obtaining stationary state, $\rho \rightarrow \rho_{\rm st}$" 
		and ``trace out" processes. 
		In the $\rho \rightarrow \rho_{\rm st}$ process, 
		we have initially 
	$\rho=\sum_{j,m,i}p_{j,m} |a_{j m i}\rangle\langle a_{j m i}|$ 
	where $	|a_{jmi} \rangle = |0\rangle \otimes |j,m,i \rangle$, 
	which is illustrated in the upper row.
	When the system becomes stationary state, we obtain
	$\rho_{\rm st} 
	=\sum_{j,m,i}\frac{p_{j,m}}{2}(|a_{j m i}\rangle 
	\langle a_{j m i}|+|b_{j m i}\rangle \langle b_{j m i}|)$ where $|b_{jmi} \rangle = |1\rangle \otimes |j,m-1,i \rangle$, and we intuitively show this in the middle row. 
	When the FQ is isolated from the $k$th qubits and initialized at the end of Step~I, 
	we obtain $\rho^{\rm I} =\sum_{j,m,i}p'_{j,m}|a_{j m i}\rangle\langle a_{j m i}|$, 
	as shown in the bottom row.  
	The gray zones represent the dark state.}
	\label{fig:a}
\end{figure}
%
%

We then cut the interaction between the 0th qubit and the others, and 
initialize the 0th qubit state to the ground state $|0\rangle $.
The final total density matrix is given as 
\begin{align}
	\label{eq:final_S_I}
	\rho^{\rm I} &=|0\rangle \langle0|\otimes{\rm Tr}_{0}(\rho_{\rm st})
	=\sum_{j,m,i}p'_{j,m}|a_{j m i}\rangle\langle a_{j m i}|,
\end{align}
where ${\rm Tr}_{0}$ denotes that only the
0th qubit
degrees of freedom is traced out.
By simple calculations, we obtain  
\begin{align}
	p'_{j,j}&=\frac{p_{j,j}}{2},\nonumber\\
	p'_{j,m}&=\frac{p_{j,m}}{2}+\frac{p_{j,m+1}}{2},~~m\neq j,-j\nonumber\\
	p'_{j,-j}&=p_{j,-j}+\frac{p_{j,-j+1}}{2}.
	\label{eq:elem}
\end{align}
Thus, the repetition of Step~I can be completely represented by the above update rule 
\eqref{eq:elem}. See Fig.~\ref{fig:a}.

\begin{figure}[t]
	\begin{center}
		\includegraphics[width=9.5cm]{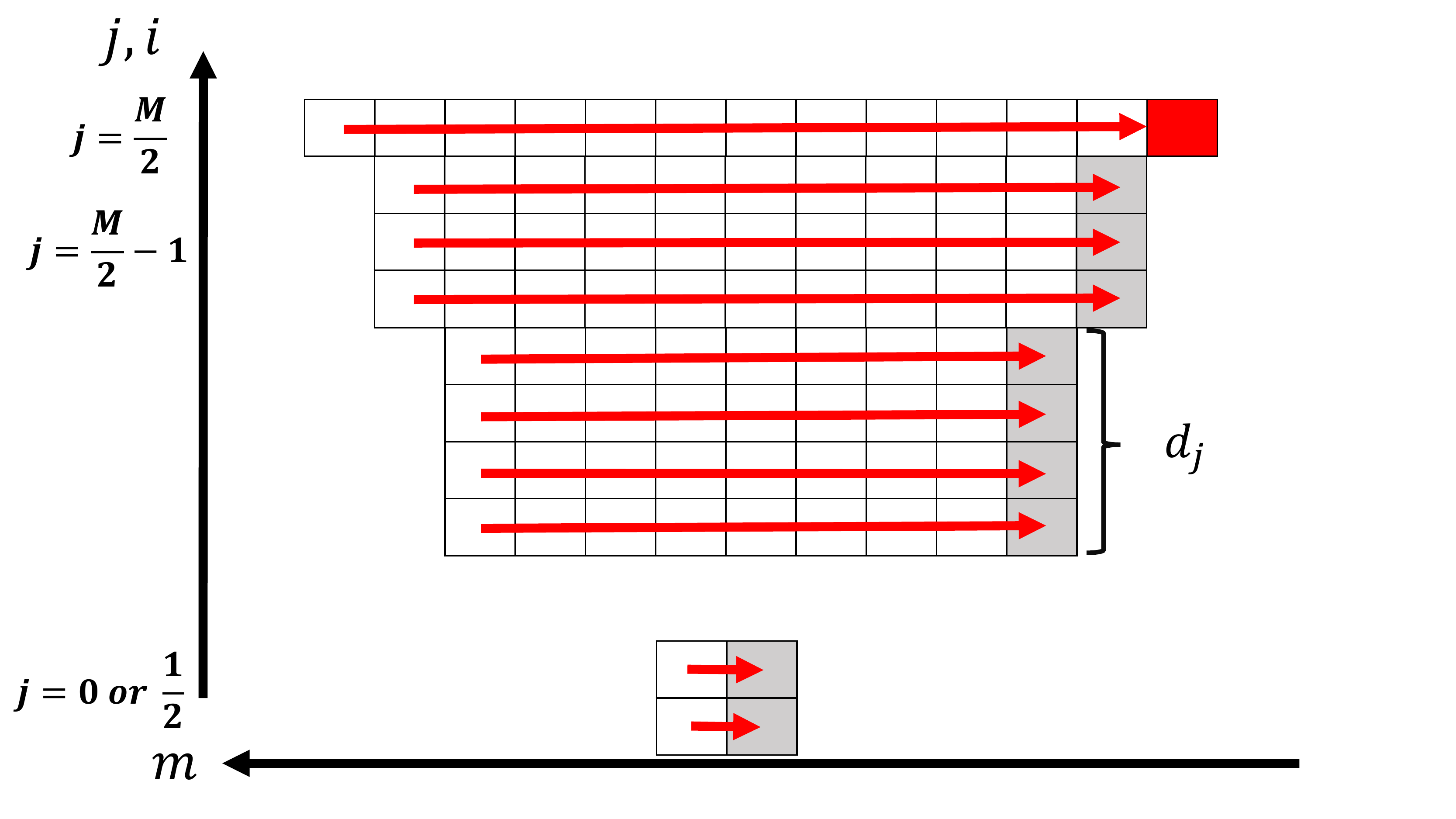}
	\end{center}
	\caption{(color online) Global view of 	the repetitive application of
	the update rule~\eqref{eq:elem} of Step~I.  
	The states of $|j,m,i\rangle$
	converges to the states of $|j,-j,i\rangle$.
	The red zone represents the target polarized state where every qubit is in the ground state.
	}
	\label{fig:g}
\end{figure}

The above processes (dynamics + trace out + initialization) is called Step~I hereinafter.
Fig.~\ref{fig:g} shows the global view of the whole density matrix dynamics 
during Step~I. 

\subsection{Step~II}
Even when we repeat the Step I, we cannot polarize all qubits, because of the existence of the so-called dark states of $|j,-j,i\rangle $ \cite{boller1991observation,fleischhauer2005electromagnetically,zhu2014observation,matsuzaki2015improving,matsuzaki2015improvinga,putz2017spectral,julsgaard2013quantum,probst2015microwave}. By the repetitive application of the step I, 
the population of each basis $|a_{jmi}\rangle\langle a_{jmi}|$ in Eq. (\ref{eq:ini}) except that of the dark states converges to zero and thus the populations will be accumulated onto these dark states.
At the infinite repetition limit, the density matrix is given as
\begin{equation}
	\rho=\sum_{j,m,i}\delta_{m,-j}p^{\infty}_{j,-j}|a_{jmi} \rangle\langle a_{jmi} |,~~p^{\infty}_{j,-j}:=\frac{2 j + 1}{2^M}. 
\end{equation}
Therefore, 
we cannot achieve perfect polarization only by Step~I. 

To overcome this problem, in Step~II,
we apply dephasing noise to the qubits ($k = 1 \sim M$) 
of which state is given by $\rho^{\rm I}$ (Eq.~\eqref{eq:final_S_I}) 
according to Eq.~\eqref{eq:lin_II},
and we wait until the dynamics converges.
We will prove that the density matrix 
after Step~II is given as
\begin{align}
	\rho^{\rm II}
	&=\sum_{j,m,i}P_{m}|a_{jmi} \rangle\langle a_{jmi}|.
	\label{eq:afterstepii}
\end{align}
The dephasing remove the $j$ dependence 
in the probability of  $|a_{jmi} \rangle\langle a_{jmi} |$.
Note that the density matrix given in Eq.~(\ref{eq:afterstepii}) is also a special case of Eq. (\ref{eq:ini}).

\begin{figure}[t]
	\begin{center}
		\includegraphics[width=9.5cm]{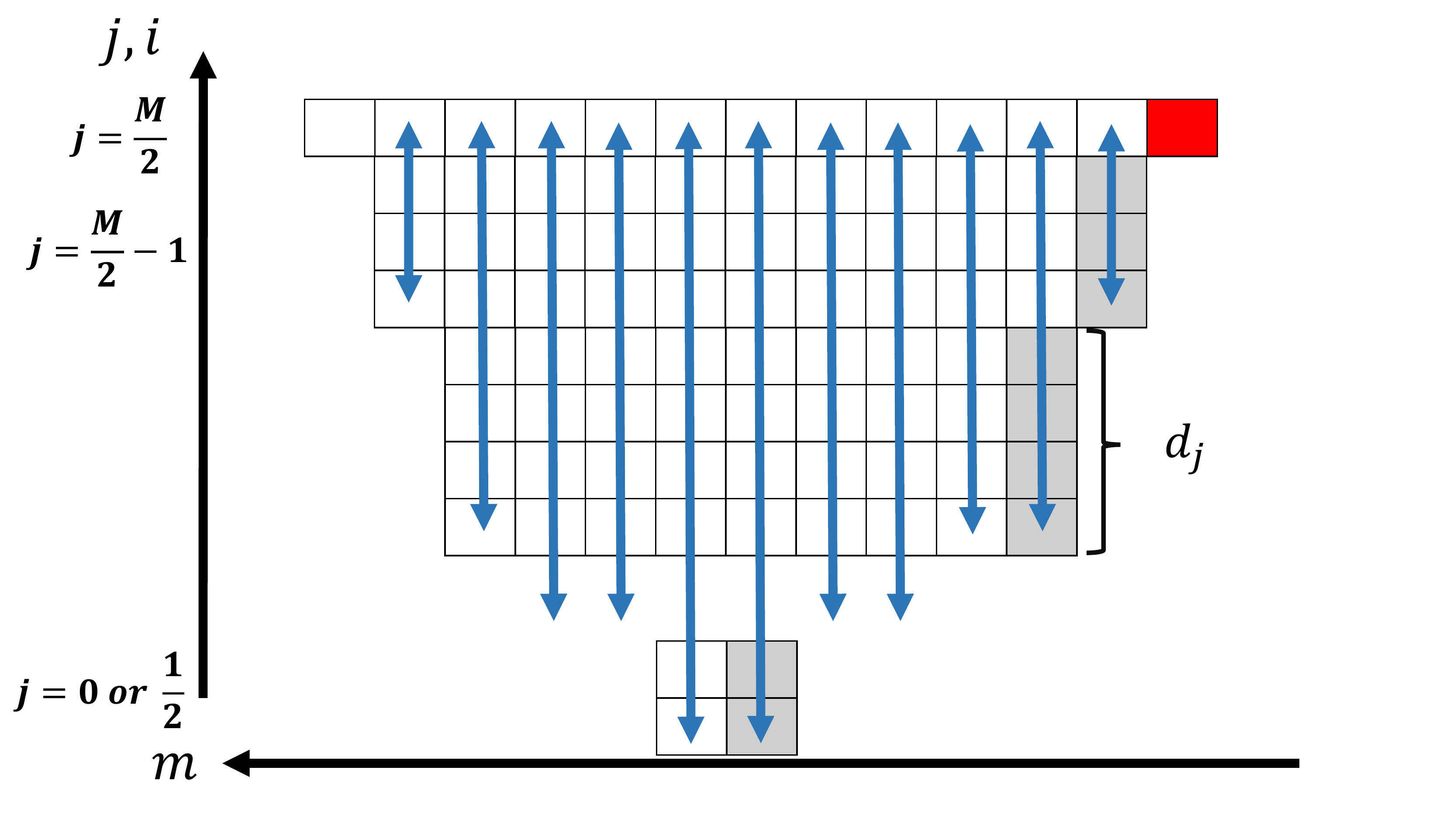}
	\end{center}
	\caption{(color online) Global view of the update
    rule~\eqref{eq:update_rule_II} of Step~II, 
    which is regarded as 
	averaging of the population of $|j,m,i\rangle$ about $j$ and $i$ for fixed $m$.
	The blue arrow in the figure denotes such an averaging.
    }
	\label{fig:h}
\end{figure}

We first introduce $\rho_{j,m}$ which is given as
\begin{equation}
	\label{eq:rhojm}
	\rho_{j,m} :=\sum^{d_{j}}_{i=1}|a_{jmi}\rangle\langle a_{jmi}|.
\end{equation}
Let us denote the operation of Step~II by ${\cal E}_{\rm II}$.
Then the dynamics of Step II is given as 
\begin{align}
    \rho^{\rm II}&={\cal E}_{\rm II}(\rho_{\rm I})=\sum_{j,m}p'_{j,m}\Bigl({\cal E}_{\rm II}(\rho_{j,m}) \Bigr).
\end{align}
The action of ${\cal E}_{\rm II}$ is written as
\begin{equation}
    {\cal E}_{\rm II}(\rho_{j,m})=\frac{d_j}{{}_MC_{m+M/2}}\sum_{s=|m|}^{M/2}\sum^{d_{s}}_{i=1}|a_{smi} \rangle\langle a_{smi} |,
    \label{eq:ActionStepII}
\end{equation}
which is shown in Appendix.
\begin{figure}[t]
	\begin{center}
		\includegraphics[width=9.cm]{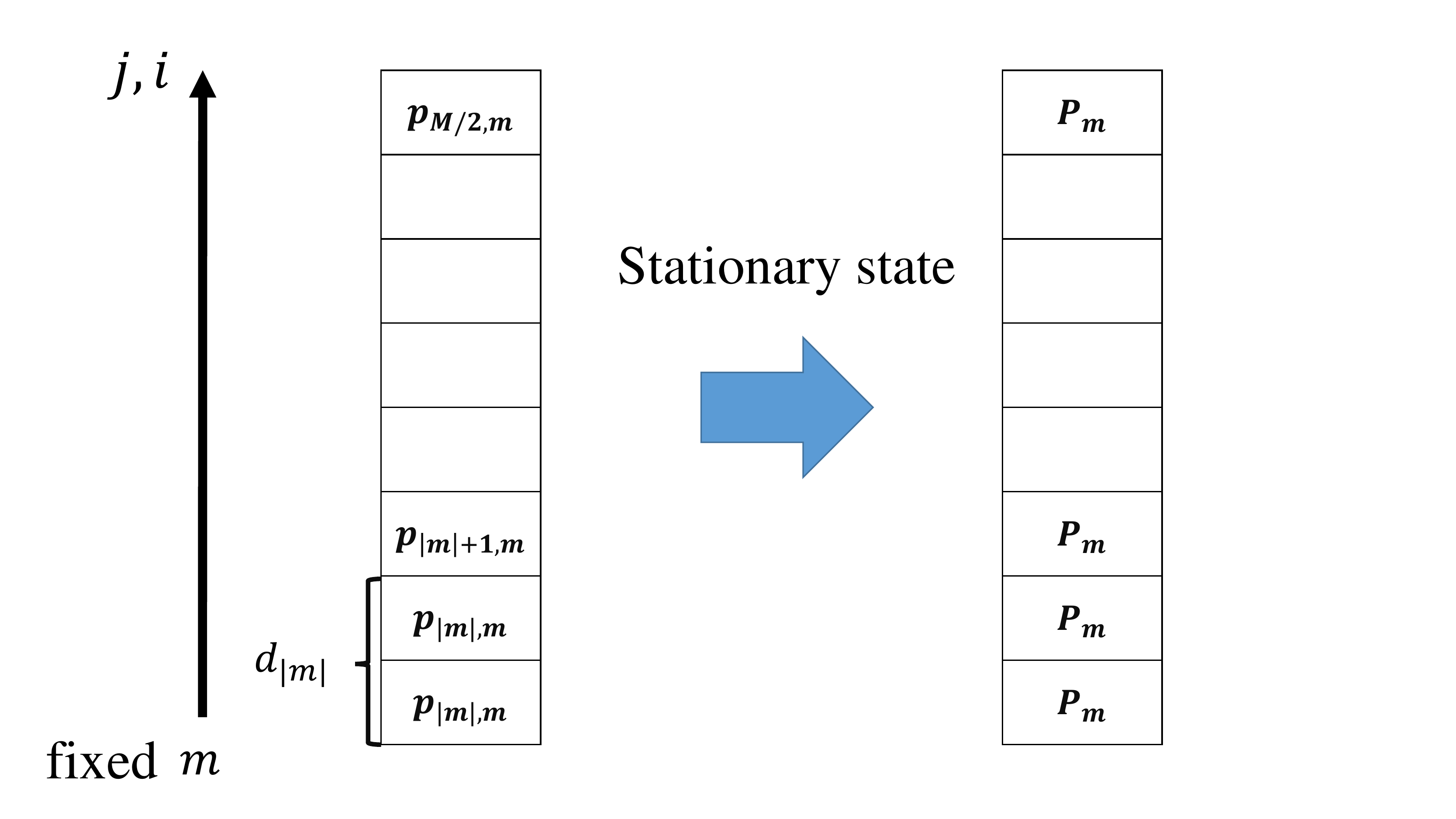}
	\end{center}
	\caption{(color online) Step~II can be regarded as a process of 
	averaging of $p_{j,m}$ about $j$ and $i$ for fixed $m$.
	After Step II, the population of $|j,m,i\rangle $ in the density matrix does not depend on $j$.
	}
	\label{fig:c}
\end{figure}
After showing the $\sum$'s in $\rho^{\rm I}$  explicitly, 
we  transform $\rho^{\rm I}$ as follows. 
\begin{align}
	& \rho^{\rm II} \nonumber \\
	&=\sum_{m=-M/2}^{M/2}\sum^{M/2}_{j=|m|}p'_{j,m}
	\left(\frac{d_j}{{}_MC_{m+M/2}} \sum_{s=|m|}^{M/2}\sum^{d_{s}}_{i=1}|a_{smi} 
	\rangle\langle a_{smi} | \right) \nonumber \\
	&=\sum_{m=-M/2}^{M/2}\left( \sum^{M/2}_{j=|m|}p'_{j,m}
	\frac{d_j}{{}_MC_{m+M/2}} \right) 
	\sum_{s=|m|}^{M/2}\sum^{d_{s}}_{i=1}|a_{smi} 
	\rangle\langle a_{smi} | \nonumber\\
	&=\sum_{m=-M/2}^{M/2}P_{m}\sum^{M/2}_{s=|m|}
	\sum^{d_{s}}_{i=1}|a_{smi}\rangle\langle a_{smi}|\nonumber\\
	&=\sum_{j,m,i}P_{m}|a_{jmi} \rangle\langle a_{jmi} |, 
\end{align}
where $P_{m}$ is given as 
\begin{align}
	P_{m}&=\sum^{M/2}_{j=|m|}p'_{j,m}\frac{d_{j}}{_{M}C_{m+\frac{M}{2}}} \nonumber \\
	&=\frac{1}{_{M}C_{m+\frac{M}{2}}}\sum^{M/2}_{j=|m|}\sum^{d_{j}}_{i=1}p'_{j,m}.
\end{align}
Because the number of orthogonal states $|a _{j m i}\rangle$ for each $m$ is $_{M}C_{m+\frac{M}{2}}$,
this process can be considered as an averaging process of $p'_{j,m}$ in $\rho^{\rm I} $ 
for a fixed $m$.
(See Fig. \ref{fig:c} for intuitive explanation of the Step~II.)
Thus, Step~II can be totally represented by the update rule,
\begin{align}
	\label{eq:update_rule_II}
	p'_{j,m} \rightarrow P_m.
\end{align}
Figure~\ref{fig:h} shows the schematic view of the whole density matrix 
dynamics after Step~II.

\subsection{Step I + Step II}

It is possible to polarize the qubits to ground states by combining Step~I and II.  
In order to consider the polarization process, 
it is convenient to employ the following variable
\begin{figure}[t]
	\begin{center}
		\includegraphics[width=9.5cm]{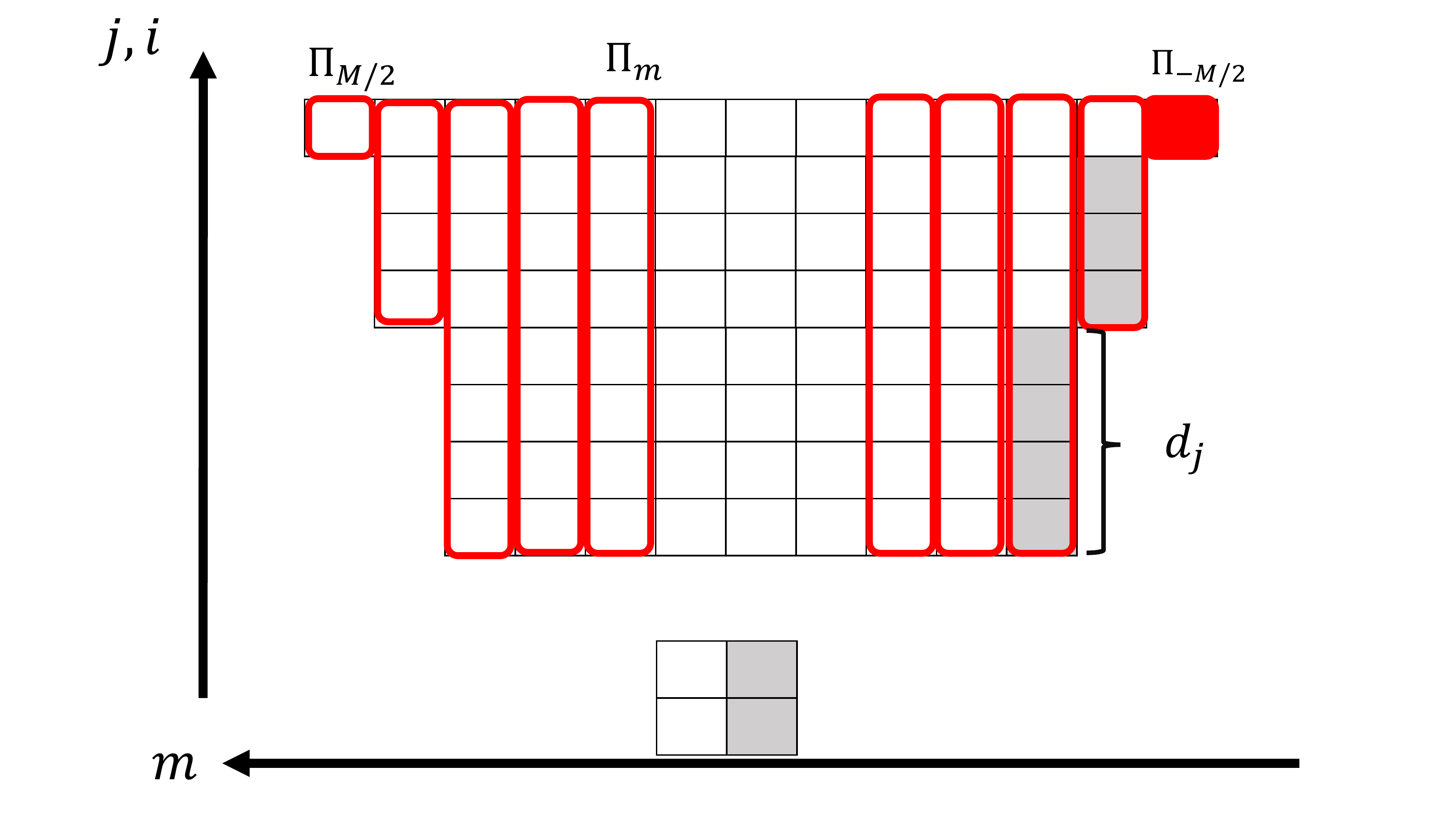}
	\end{center}
	\caption{(color online) Definition of $\Pi_{m}$
	where we sum up all populations of $|j,m,i\rangle $ in the density matrix
	for a fixed $m$.
	}
	\label{fig:d}
\end{figure}
\begin{equation}
	\label{eq:pro_col_m}
	\Pi_{m}=\sum^{M/2}_{j=|m|}\sum^{d_{j}}_{i=1}p_{j,m},
\end{equation}
which is the total probability of the $m$-th column states, see Fig.~\ref{fig:d}.
These states have the same energy.
If $p_{j,m}=P_{m}$ (after Step~II and the initial state), this is given as 
\begin{equation}
	\Pi_{m}= {}_{M}C_{m+\frac{M}{2}} P_{m}.
\end{equation}

We, now, consider the update rule of $\Pi_{m}$ under Step~I and II.
Let us denote the total probability of the $m$-th column states after the $n$ repetition by $\Pi^{(n)}_{m}$.
The update rule depends on the sign of $m$.

\begin{figure}[b]
	\begin{center}
		\includegraphics[width=9.5cm]{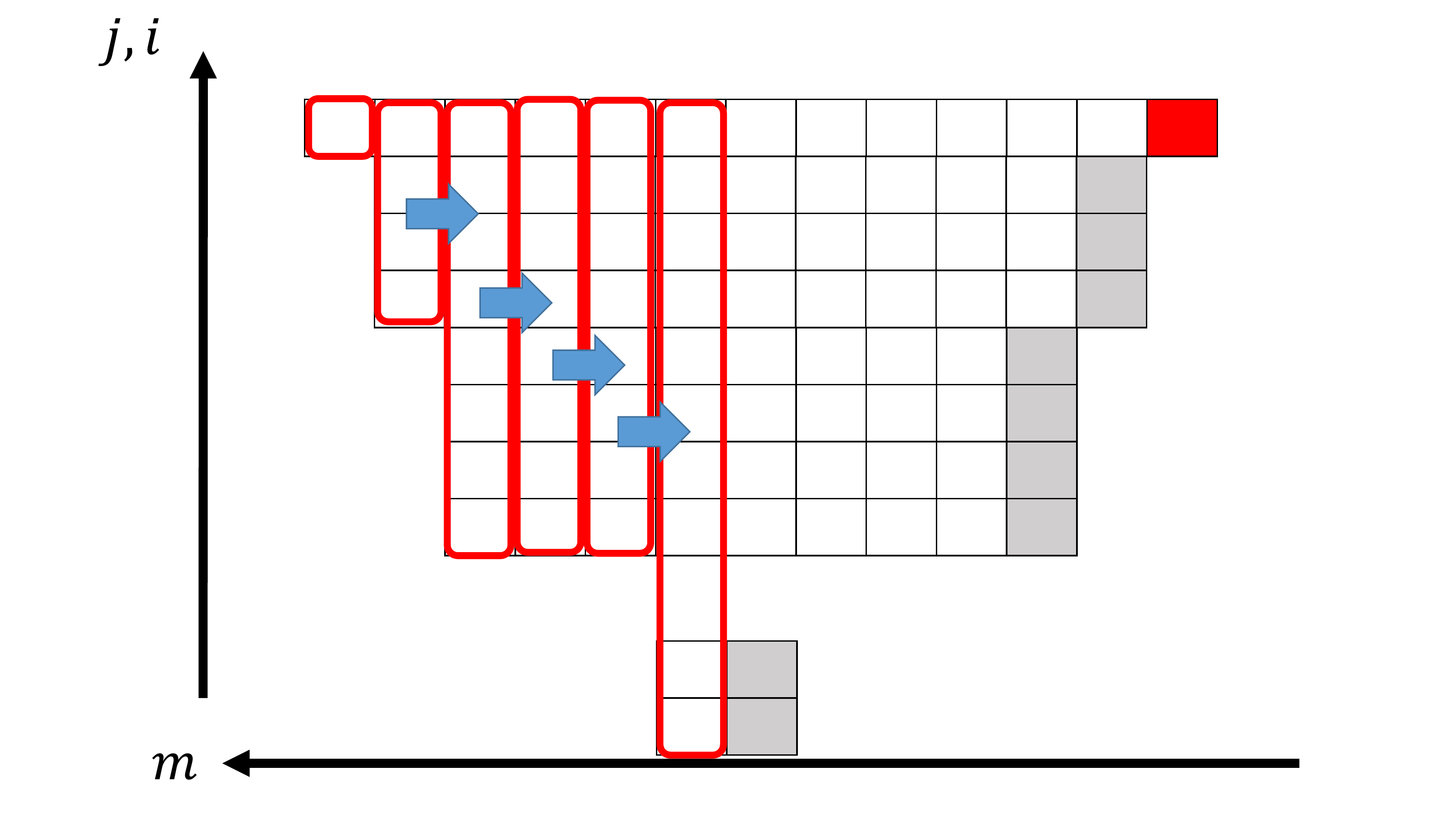}
	\end{center}
	\caption{(color online) Update rule for $m>0$.
	Half of the population of $|j,m,i\rangle $ is transferred to that of $|j,m-1,i\rangle $.
	}
	\label{fig:e}
\end{figure}

First,
we consider the case of positive $m$.
These columns have no dark state,  as shown in Fig.~\ref{fig:e}.
According to the update rule (\ref{eq:elem}), all elements in the $m$-th column can give half of its probability $P_{m}$ 
to the right ones and get half of the probability from the left ones by Step~I, 
as shown in Fig. \ref{fig:e}. Thus the update rule is given as 
\begin{align}
	\Pi^{(n)}_{M/2}&=\Pi^{(n-1)}_{M/2}/2, \\
	\Pi^{(n)}_{m}&=\Pi^{(n-1)}_{m}/2+\Pi^{(n-1)}_{m+1}/2,
	~~{\rm for}~~m>0,~m\neq M/2.
	\nonumber 
\end{align}

\begin{figure}[t]
	\begin{center}
		\includegraphics[width=9.5cm]{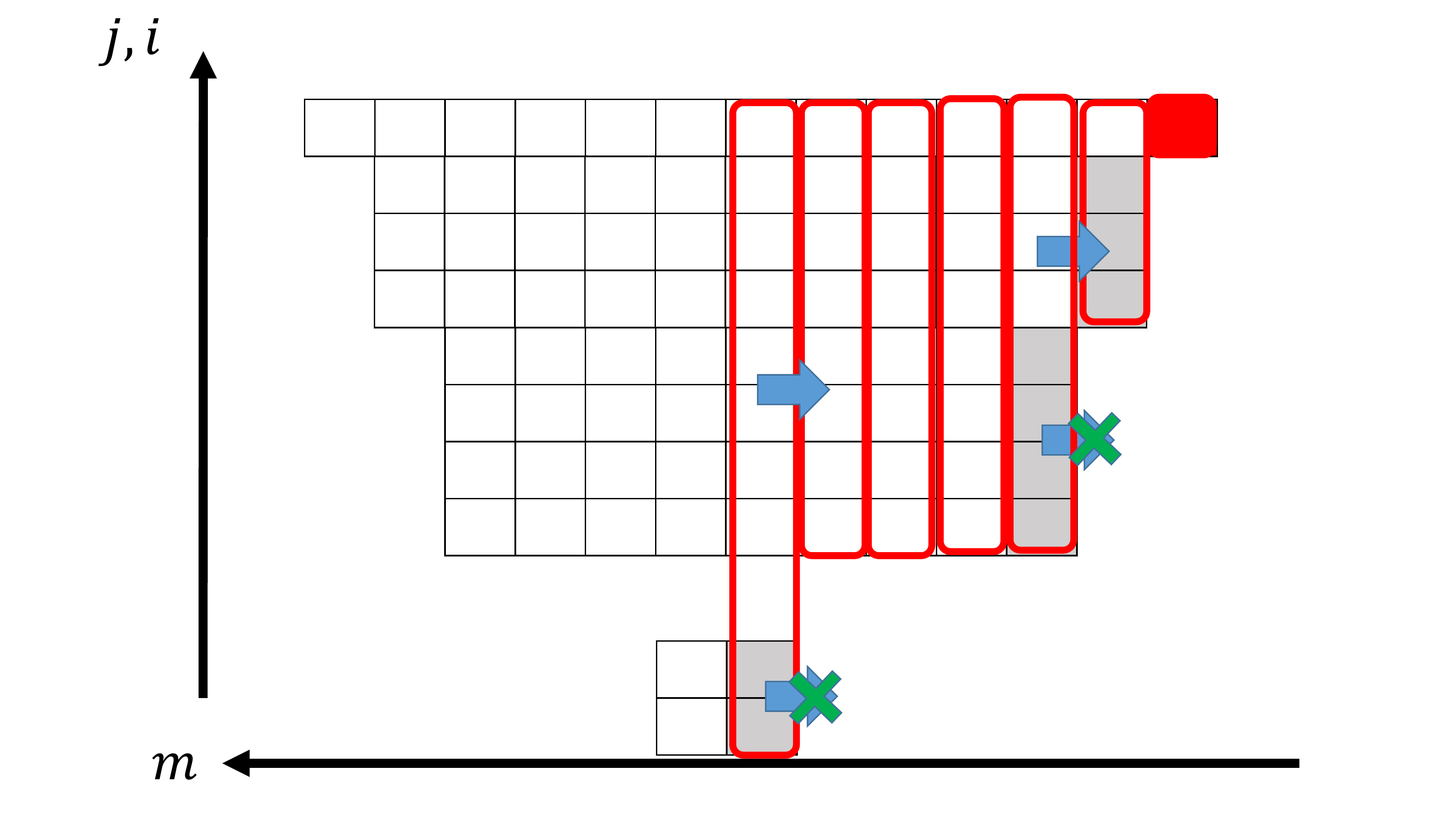}
	\end{center}
	\caption{(color online) Update rule for $m \le 0$. 
	Half of the population of $|j,m,i\rangle $ is transferred to that of $|j,m-1,i\rangle $ unless $|j,m,i\rangle $ is a dark state that cannot transfer its population by the Step I.
	}
	\label{fig:f}
\end{figure}

Second, we consider the case for $m\le 0$ where we have
$d_{|m|}$ dark states (see Fig. \ref{fig:f}).
Then, only $({}_{M}C_{m+\frac{M}{2}}-d_{|m|})$ elements can give the half of its probability 
$P_{m}$ to the right ones.
Because each element in the $m$-th column have the same probability $P_{m}$ due to Step~II, we only have to 
count how many dark states in the $m$-th column in order to derive the update rule.
A ratio between the number of the dark states and that of the total elements 
for a fixed $m$
determines the amount of the population to be transferred from the $m$-th column to the $(m+1)$-th column (see Fig. \ref{fig:f}).
Thus, the update rule for the $m\leq 0$ case is given as 
\begin{align}
	\Pi^{(n)}_{m}&=\frac{1}{2}\Bigl(1+\frac{d_{|m|}}{_{M}C_{m+\frac{M}{2}}}\Bigr)\Pi^{(n-1)}_{m}
	\nonumber \\
	&+\frac{1}{2}\Bigl(1-\frac{d_{|m+1|}}{_{M}C_{1+m+\frac{M}{2}}}\Bigr)\Pi^{(n-1)}_{1+m},
	\nonumber \\
	&~~~~~~~~~~{\rm for}~~m\leq 0,~m\neq-M/2,\nonumber\\
	\Pi^{(n)}_{-M/2}&=\Pi^{(n-1)}_{-M/2}
	+\frac{1}{2}\Bigl(1-\frac{d_{|1-M/2|}}{_{M}C_{1}}\Bigr)\Pi^{(n-1)}_{1-M/2}.
\end{align}
The probability of each element $p_{j,m}$ in the $m$-th column can have different values 
after Step~I because the rows labeled by $(j,i)$ are not equivalent. 
For instance, the probabilities of dark states will be relatively large comparing with those of 
the other states, see Fig. \ref{fig:f}.
If the probabilities $p_{j,m}$ depends on $j$, the above update rule does not 
work because this rule is derived under the assumption that $p_{j,m}=P_{m}$.
Thus, when we repeat only Step I, the variable $\Pi^{}_{m}$ is not appropriate to describe our process and we should use the original update rule (\ref{eq:elem}) for $p_{j,m}$.
On the other hand, when we insert Step~II, since this step averages these probabilities, this update rule for $\Pi_{m}$ can be applied  for the next repetition of Step~I.

Let us summarize the combined process of Step~I and II. 
Note that the state after $n$ times repetition is given as
\begin{align}
	 \rho^{(n)} 
	 &= \sum^{M/2}_{m=-M/2}P^{(n)}_{m}\sum^{M/2}_{j=|m|} 
	 \sum^{d_{j}}_{i=1}|a_{j,m,i} \rangle\langle a_{j,m,i} |, 
\end{align}
which is justified with the calculation above.
By introducing the probability of the $m$-th column after $n$ times repetition
like Eq.~\eqref{eq:pro_col_m},
\[
\Pi^{(n)}_{m}
=\sum^{M/2}_{j=|m|}\sum^{d_{j}}_{i=1}P^{(n)}_{m}
={}_{M}C_{m+\frac{M}{2}}P^{(n)}_{m},
\]
the update rule is summarized as
\begin{align}
	\Pi^{(n+1)}_{M/2}&=\Pi^{(n)}_{M/2}/2,\nonumber\\
	\Pi^{(n+1)}_{m}&=\Pi^{(n)}_{m}/2+\Pi^{(n)}_{m+1}/2,~~{\rm for }~~m>0,~m\neq M/2,\nonumber\\
	\Pi^{(n+1)}_{m}&=\frac{1}{2}\Bigl(1+\frac{d_{|m|}}{_{M}C_{m+\frac{M}{2}}}\Bigr)\Pi^{(n)}_{m} 
	\nonumber \\
	&+\frac{1}{2}\Bigl(1-\frac{d_{|m+1|}}{_{M}C_{1+m+\frac{M}{2}}}\Bigr)\Pi^{(n)}_{1+m},
	\nonumber \\
	&~~~~~~~~~~~~{\rm for }~~m\leq 0,~m\neq-M/2, \nonumber\\
	\Pi^{(n+1)}_{-M/2}&=\Pi^{(n)}_{-M/2}+\frac{1}{2}\Bigl(1-\frac{d_{|1-M/2|}}{_{M}C_{1}}\Bigr)
	\Pi^{(n)}_{1-M/2}.
	\label{eq:update}
\end{align}
As discussed previously, this update rule is based on the fact that $P^{(n)}_{m}$ 
is independent of $i$ and $j$ thanks to Step II.
By using above $\Pi^{(n)}_{m}$, the density matrix after $n$-times repetition is given by
\begin{align}
	 \rho^{(n)} 
	 &= \sum^{M/2}_{m=-M/2}\frac{\Pi^{(n)}_{m}}{{}_{M}C_{m+\frac{M}{2}}}\sum^{M/2}_{j=|m|} 
	 \sum^{d_{j}}_{i=1}|a_{j,m,i} \rangle\langle a_{j,m,i} |.
\end{align}
%

\begin{figure}[t]
	\begin{center}
		\includegraphics[width=8.0cm]{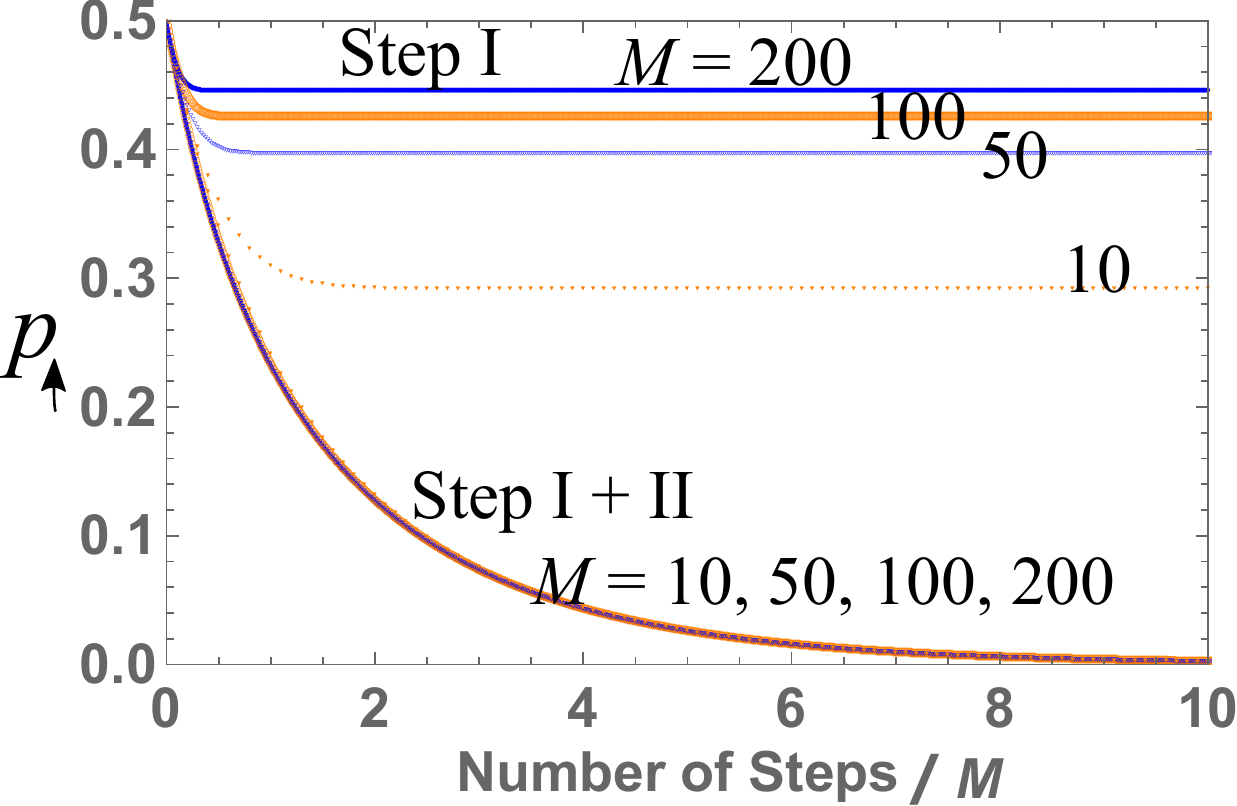}
	\end{center}
	\caption{(color online) 
		Plot of the excited-state population, $p_{\uparrow}$, of the qubit
		against the normalized number of steps. 
		When we adopt only Step~I,  $p_{\uparrow}$ 
		converges to finite non-zero values. On the other hand, when we adopt 
		the Step~I and II, $p_{\uparrow}$  
		approaches to zero, and the plots behave same regardless of the number of qubits.
		Here, we take $M=10, 50,100$ and 200
		from the bottom to top where $M$ denotes the number of qubits. 
	}
	\label{fig:i}
\end{figure}

Let us consider a probability of an excited state of the $k$-th spin,
\begin{equation}
    p_{\uparrow, k}=\frac{1}{2}\bigl(1+{\rm Tr}(\sigma^{(k)}_{z}\rho)\bigr).
\end{equation}
Note that all spins are now equivalent.
In this case, $p_{\uparrow, k}$
has no dependency on $k$, and thus we drop the index $k$ in this section.
The dynamics of $p_{\uparrow}$ is summarized in Fig.~\ref{fig:i}
when $M=10, 50, 100$ and $200$. The $x$-axis is 
the number of steps divided by $M$. 
Since we assume that we wait until the system saturates at each Step~I and II, the plot is 
independent of
the interaction strength $g$ in Eq.~\eqref{eq:ham_I} and 
the dephasing rate of the spins $\gamma'$ in  Eq.~\eqref{eq:lin_II}.
When we perform only Step I, the population will be trapped by the dark states, 
and the final population of the excited state of the spins increases as $M$ does. 
On the other hand, the excited-state population converges to zero 
when we perform both Step~I and II, as expected. Also, the plot of Step~I+II shows 
a universal dynamics and does not depend on $M$.

\section{
	Spin polarization with a flux qubit}
\label{sec:NS}

We analyze a realistic polarization dynamics of electron spins 
based on the discussion in \S~\ref{sec:theory} and show 
numerical simulations in various conditions. 
We consider the $0$th qubit as the flux qubit (FQ), which is 
highly controllable, and
and consider the other qubits as the electron spins. 

\subsection{Parameters}
\label{subsec:parameters}
\begin{figure}[b]
	\begin{center}
	\includegraphics[width=8.0cm]{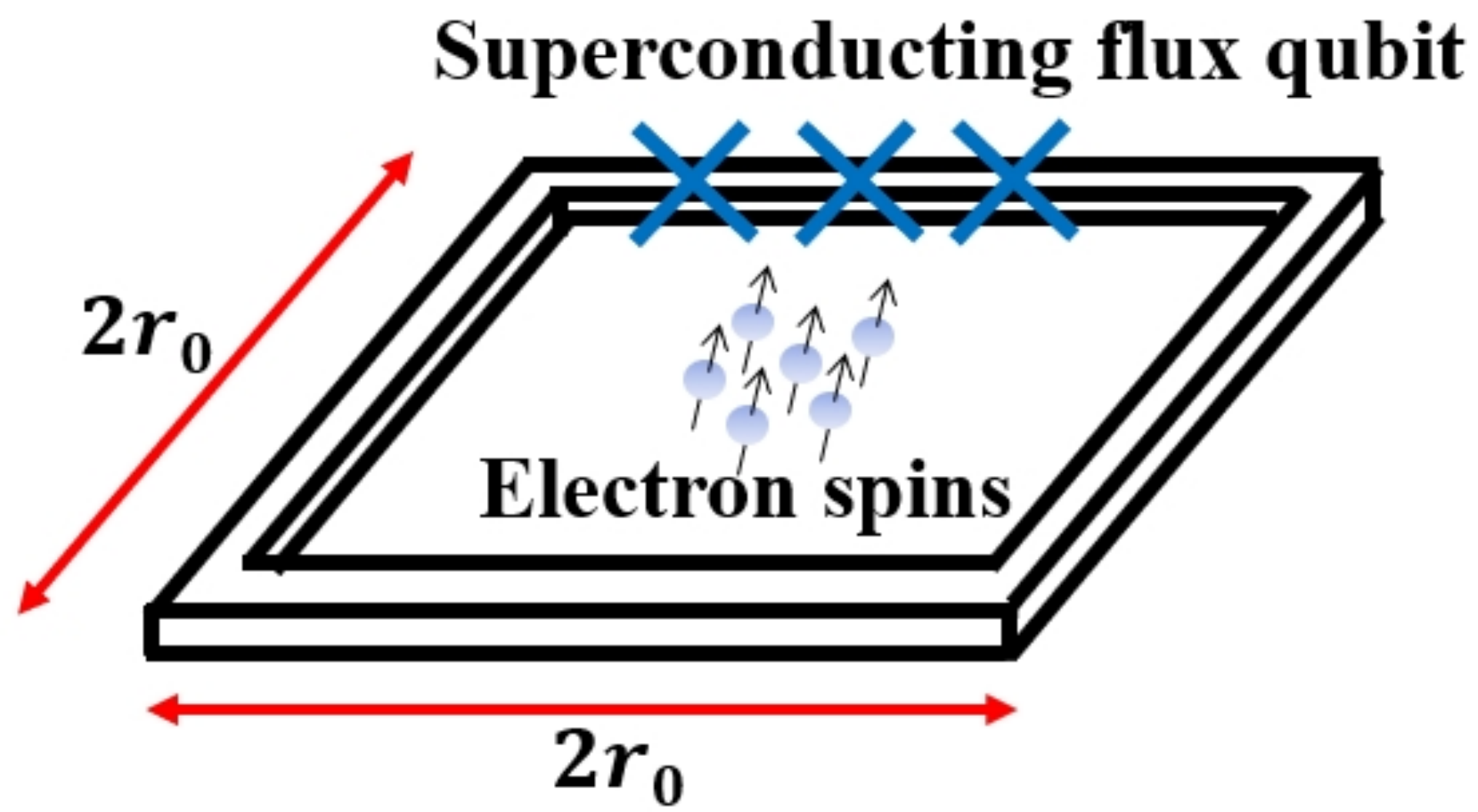}
	\end{center}
	\caption{
	(color online)  Illustration of our system composed of the superconducting flux qubit (FQ) 
	and electron spins. The FQ consists in a square loop containing three Josephson junctions,
	illustrated as three blue $\times$'s.
	The electron spins are located in the middle of the FQ, and they are inductively coupled with the FQ.}
	\label{fig:fqspin}
\end{figure}

We simulate the polarization dynamics with realistic parameters according to the 
Hamiltonian~\eqref{eq:ham_FQ_e} and Lindbladian~\eqref{eq:lin_FQ_e}. 
The configuration of the FQ is assumed to be a $2r_0 \times 2r_0$ square
(see Fig. \ref{fig:fqspin})
and the parameters for numerical simulations are summarized in Table~\ref{tab:FQ_p}. 
Since fast reset of the superconducting qubit has been demonstrated with a resetting time of $120$ ns
where the the longitudinal relaxation time $T_1$ of the qubit can be controlled over a factor of $50$
\cite{reed2010fast,TunableRefrig2020}, we assume that the time required for initialization is 
$t_{\rm i} = 5 \times 10^{-6}$~s. 

The electron spins are located in the middle of a square
determined by the FQ (see Fig. \ref{fig:fqspin}). 
Their interaction strengths $g_k$'s  with the
FQ during the spin lock
are given as
\cite{marcos2010coupling,twamley2010superconducting,matsuzaki2015improving}
\begin{align}
	\label{eq:gk}
	g_k &= \frac{\varepsilon}{\sqrt{\varepsilon^2 + \Delta^2} }
	\frac{\gamma_e \mu_0 I_p}{2 \pi}
	\sum_{i=1} ^4\frac{1}{r_i^{(k)}} ,
\end{align}
where $r_i^{(k)}$, $\gamma_e$, $\mu_0$ are the distance from $i$-th side 
of the FQ to an $k$-th spin, the gyromagnetic ratio of an electron spin, and 
the Vacuum permeability, respectively. If the electron spin is placed in the middle 
of a FQ, we obtain $g_k =g_0= 175$~rad/s with the parameters given in 
Table~\ref{tab:FQ_p}. This value gives the energy scale of the interaction between 
the FQ and spins.

\begin{table}[h]
	\begin{tabular}{|c|c|c|}
		\hline 
		parameter of FQ & symbol &value \\
		\hline \hline 
		size & $r_0$ & $3.0 \times 10^{-6}$~m \\
		persistent current & $I_p$ & 180~nA \\
		longitudinal relaxation time &$T_1^{\rm (FQ)}$ & $200 \times 10^{-6}$~s \\
		transversal relaxation time & $T_2^{\rm (FQ)}$ & $30 \times 10^{-6}$~s \\
		energy gap & $\Delta/2\pi$ & $5.37 \times 10^9$~Hz \\
		detuning parameter & $\varepsilon /2\pi$ &   $0.112 \times 10^9$~Hz \\
		time required for initialization & $t_{\rm i}$ & $5 \times 10^{-6}$~s \\
		interval between initialization & $t_{\rm int}$ & $95 \times 10^{-6}$~s \\
		\hline
	\end{tabular}
\caption{Parameters of the FQ for numerical simulations \cite{bylander2011noise,yan2016flux}.}
\label{tab:FQ_p}
\end{table}

We also assume that the longitudinal relaxation time $T_1^{(e)} = 1.0$~s and the transversal relaxation time $T_2^{(e)}  = 1.0 \times 10^{-3}$~s 
for electron spins \cite{tyryshkin2012electron,herbschleb2019ultra,amsuss2011cavity}.

\subsection{Simulations}
\label{subsec_sim}
We numerically calculate the system dynamics according to the operator sum formalism
\cite{zanardi1998dissipation,kondo2016using,bando2020concatenated}.
The density matrix $\rho$ is updated as follows. 
\begin{align}
	\rho (t) \rightarrow  \rho' =e^{-i {\cal H} \delta} \rho  \, e^{i {\cal H} \delta}
	\rightarrow \rho(t+\delta) = \rho' +{\cal L}[\rho'] \delta , 
\end{align}
where we take $\delta = 5 \times 10^{-6}$~s which is small enough 
compared with the characteristic time scale such as $T_1, T_2$ of 
the FQ or spins, and $1/g_k$.  
We let the system evolve by this formalism for a time $t_{\rm int}$.
We consider various ${\cal H}$'s and  ${\cal L}$'s 
by changing parameters in \S~\ref{subsec_sim} 1 $\sim$ 4.  

The initial state of the electron spin is a completely mixed state. 
We assume that the FQ is periodically initialized into the ground state
at $t_n=(n-1)(t_{\rm i} + t_{\rm int}$) where $n$ denotes natural numbers.
We define this period as a single step.
Since the initialization of the FQ can be much faster than the time scale of the decay of the electron spins, we assume that the state of the electron spins does not change during the initialization of
the FQ.

In the numerical calculations, we do not separate the dynamics into Step~I and Step~II.
By applying both
the dephasing of the electron spins and the interaction between the FQ and the electron spins,
we simultaneously perform Step~I and Step~II.

\subsubsection{$\omega _k' = 0$, \mbox{$g_k = g_0$},  
	and $\gamma_{\rm T}^{\rm (l)} = \gamma_{\rm L}^{\rm (l)}= 0$ case}
\begin{figure}[h]
	\begin{center}
		\includegraphics[width=8.0cm]{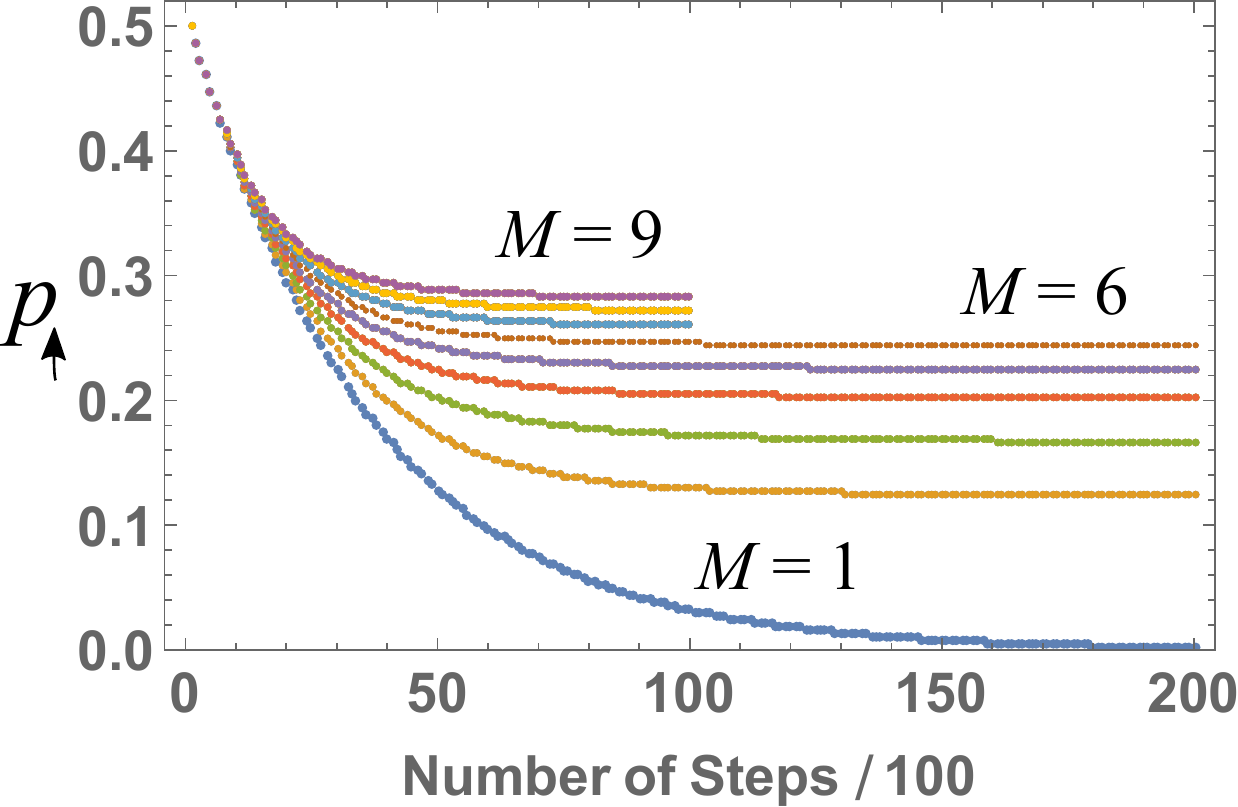}
	\end{center}
	\caption{(color online)
	Plot of the excited-state population, $p_{\uparrow}$,  of the spins
	against the number of steps for $M = 1 \sim 9$ 
	of electron spins. The initial state of the
	spins is the completely mixed state. We set the parameters as
	$\omega_{k= 1 \sim M}' = 0$, $ g_{k= 1 \sim M} =g_0 $, and
	$\gamma_{\rm T}^{\rm (l= 0 \sim M)} = \gamma_{\rm L}^{\rm (l= 0 \sim M)}= 0$. 	
	}
	\label{fig:sim_1}
\end{figure}

We first simulate the case when there is no decoherence in order to illustrate 
the influence of dark states on the polarization process. 
We assume that ${\omega _k}' = 0$ and \mbox{$g_{k=1\sim M} = g_0$}  
in Eq.~\eqref{eq:ham_FQ_e} and 
$\gamma_{\rm T}^{\rm (l=0\sim M)} = \gamma_{\rm L}^{\rm (l=0 \sim M)}= 0$ 
in Eq.~\eqref{eq:lin_FQ_e}. 
Figure~\ref{fig:sim_1} shows the dynamics of $p_{\uparrow}$. 
Note that all spins are equivalent and thus 
$p_{\uparrow, k}$'s are identical.  
Because of the dark states, $p_{\uparrow}$ saturates
in the large step limit.
Note also that the cooling rate 
in this simulation is much slower 
than
that shown in Fig.~\ref{fig:i}.  
This is because, in Fig.~\ref{fig:sim_1},
the interaction time $t_{\rm{int}}$ is much smaller than $1/g_k$, and the population transfer between the electron spins and the FQ is small at a single step.  On the other hand, in Fig.~\ref{fig:i}, half of the ground-state population is transferred to the electron spins at a single step.

\subsubsection{$\omega _k' = 0$ and 
	$\gamma_{\rm T}^{\rm (l)} = \gamma_{\rm L}^{\rm (l)}= 0$ case}
We consider the case when $g_k$ is 
inhomogeneous
because of random spin positioning 
on a substrate according to Eq.~\eqref{eq:gk}. 

\begin{figure}[h]
	\begin{center}
		\includegraphics[width=8.0cm]{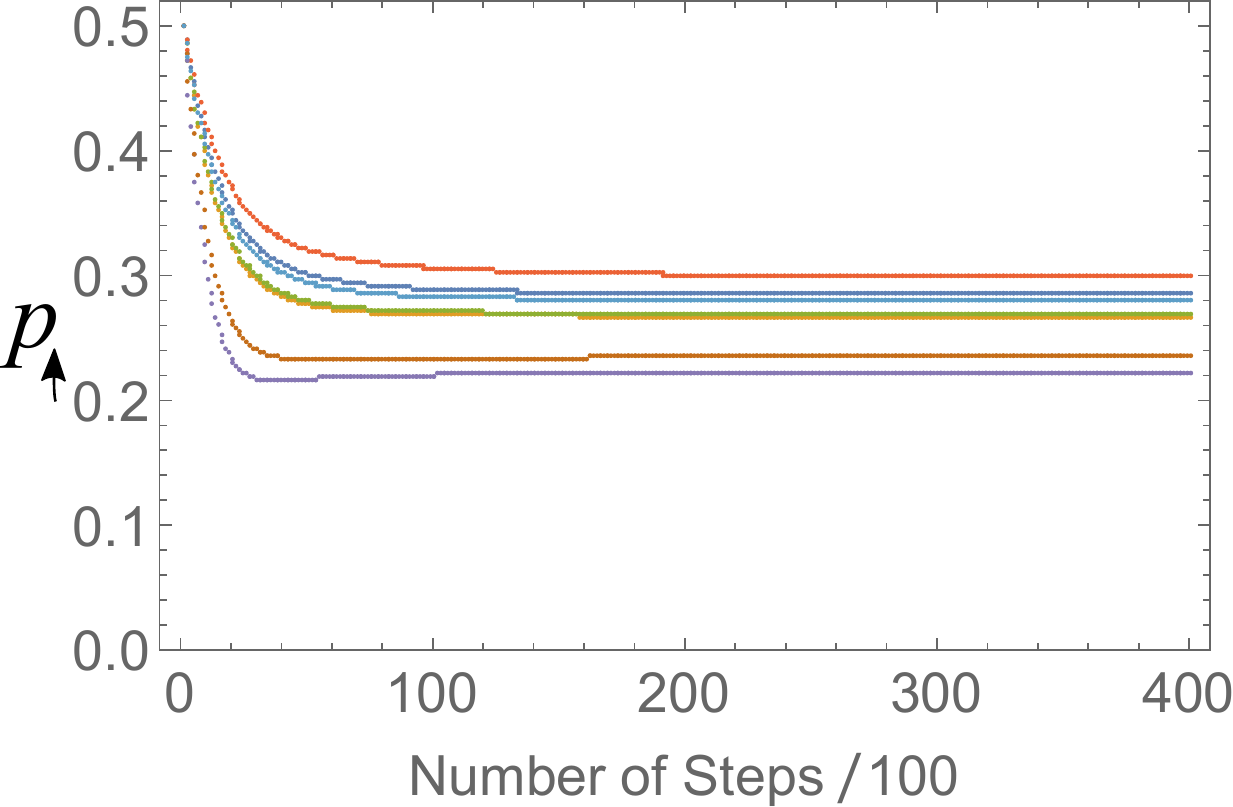}
	\end{center} 	
	\caption{(color online) 
    Plot of the excited-state population, $p_{\uparrow, k}$,  of the spins
	against the number of steps for $M=7$ when 
		$\omega_{k= 1 \sim M}' = 0$, $ \{g_{k= 1 \sim M}\} = \{ 151, 221, 173, 204, 197, 176, 180\}$~rad/s, 
		$\gamma_{\rm T}^{\rm (l= 0 \sim M)} = \gamma_{\rm L}^{\rm (l= 0 \sim M)}= 0$. 	The different lines
	correspond to $g_k$. 	}
	\label{fig:sim_2}
\end{figure}

Figure~\ref{fig:sim_2} shows $p_{\uparrow,k}$ when $M= 7$. 
Due to the different values of $g_k$ (see the caption of Fig.~\ref{fig:sim_2}), 
$p_{\uparrow,k} $ approaches a different saturation value
and does not approach zero.

\subsubsection{$\omega _k' = 0, \gamma_{\rm T}^{\rm (l)} \ne 0$, and 
	$ \gamma_{\rm L}^{\rm (l)}= 0$ case}
\begin{figure}[h]
	\begin{center}
		\includegraphics[width=8.0cm]{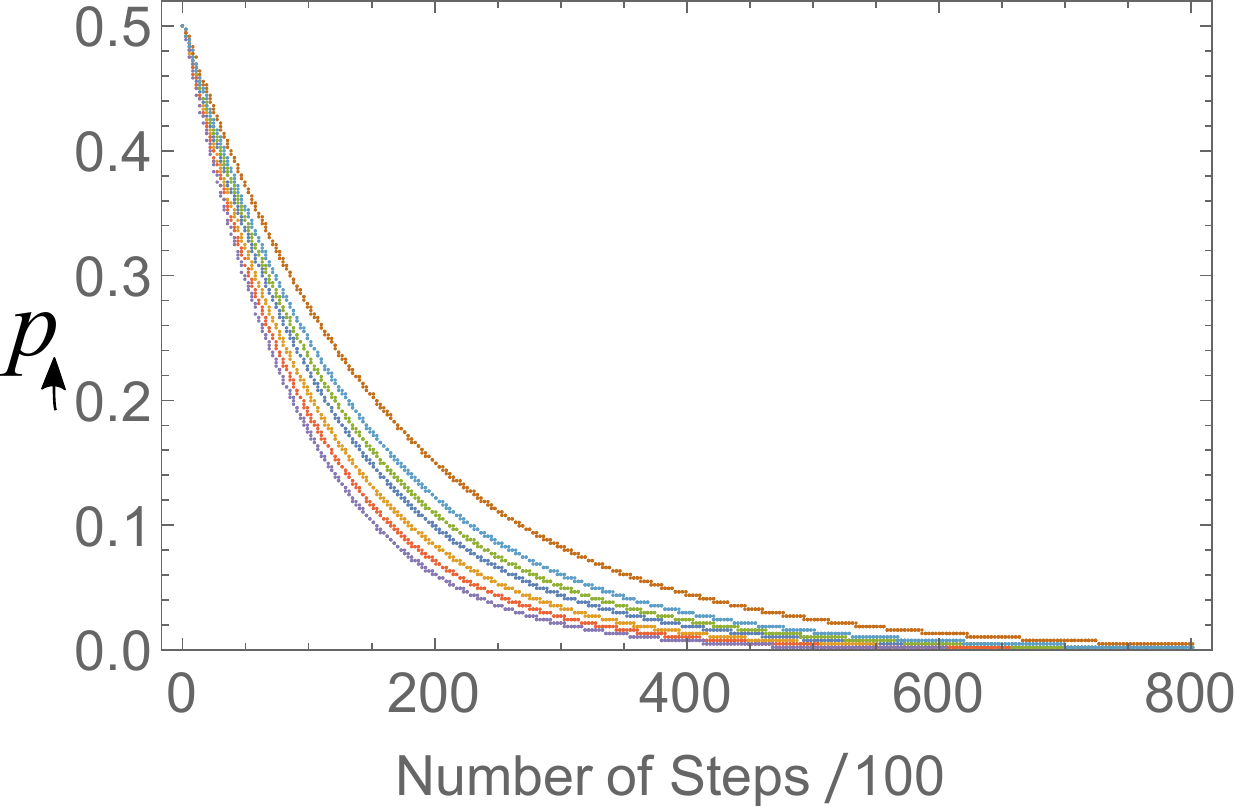}
	\end{center}
	\caption{(color online) 
	Plot of the excited-state population, $p_{\uparrow, k}$,  of the electron spins
	against the number of steps for $M=7$ when 
		$\omega_{k= 1 \sim M}' =  0$, 
		$ \{g_{k= 1 \sim M}\} = \{179, 202, 194, 161, 178, 204, 156 \}$~rad/s, 
		$\gamma_{\rm L}^{\rm (l= 0 \sim M)}= 0$, $\gamma_{\rm T}^{(0)}=1/T_2^{\rm (FQ)}$, 
		and $\gamma_{\rm T}^{(k=1 \sim M)}= 1/T_2^{(e)}$. }
	\label{fig:sim_3}
\end{figure}

We consider the case when $g_k$ is 
inhomogeneous with a finite dephasing rate of
$\gamma_{\rm T}^{\rm (l=0 \sim M)} \ne 0$. 
 Figure~\ref{fig:sim_3} shows the case with $M=7$. Due to the dephasing on spins, 
$p_{\uparrow,k}$  approaches zero with different cooling rate according to $g_k$. 
This observation with Fig.~\ref{fig:sim_2} shows that the dephasing leads $p_{\uparrow, k} =0$
at the large step limit. 

\subsubsection{realistic case} 
To support our simplified model discussed in \S~\ref{sec:theory},
we have considered non-realistic cases in \S~\ref{subsec_sim}~$1 \sim 3$ 
where some imperfections have been ignored.
We will, here, discuss a realistic case when both $\omega_k'$ and $g_k$ 
are inhomogeneous
with finite decay rates of
$\gamma_{\rm T}^{\rm (l)} \ne 0$ and $ \gamma_{\rm L}^{\rm (l)}\ne 0$.  

\begin{figure}[h]
	\begin{center}
		\includegraphics[width=8.0cm]{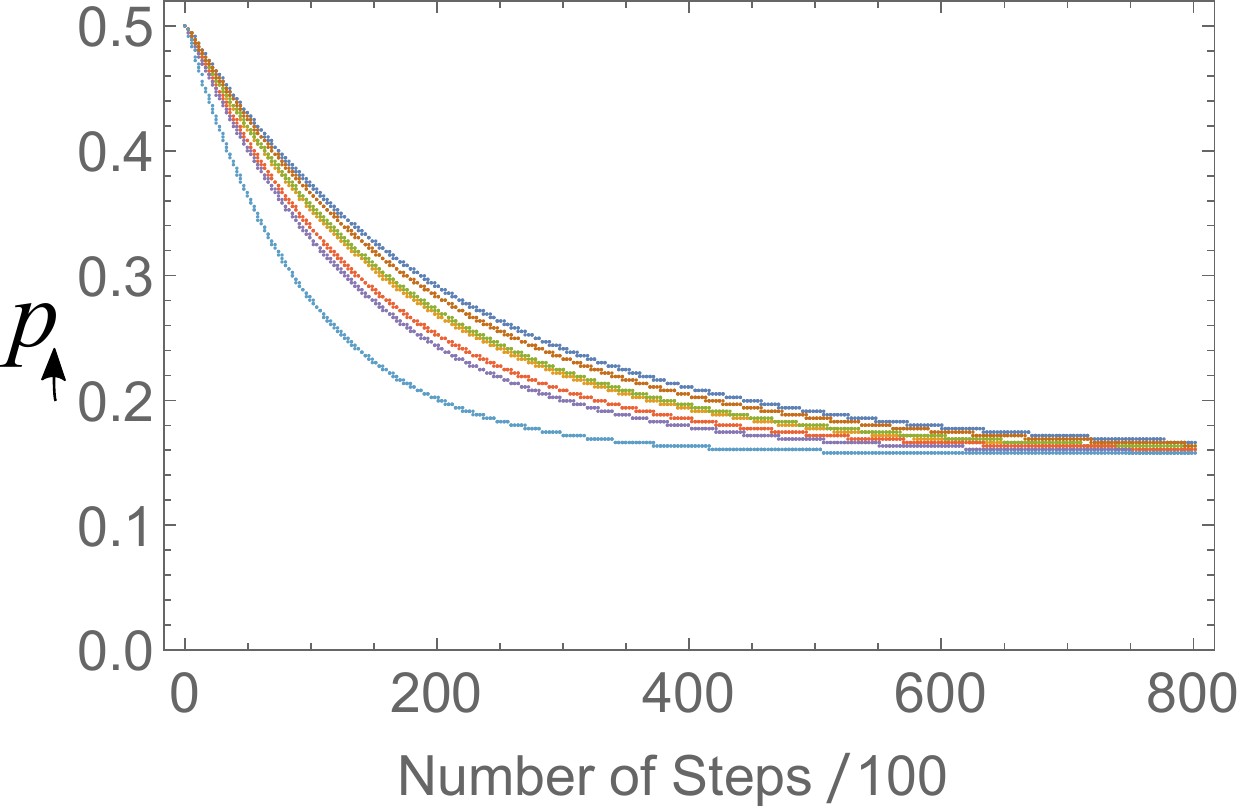}
	\end{center}
	\caption{(color online) Realistic simulations.
		We plot the excited-state population, $p_{\uparrow, k}$,   of the electron spins
		against the number of steps	for $M=7$ when 
		$\omega_{k= 1 \sim M}' =   \{ 4736, 455, -6867, 1773, -1569, 703, -5204\}$~rad/s, 
		$\{g_{k= 1 \sim M}\} = \{ 193, 163, 175, 225, 178, 160, 268 \}$~rad/s, 
		$\gamma_{\rm L}^{(0)} = 1/T_1^{\rm (FQ)}$,  $\gamma_{\rm L}^{({k= 1 \sim M})} = 1/T_1^{(e)}$, 
		$\gamma_{\rm T}^{(0)}=1/T_2^{\rm (FQ)}$, 
		and $\gamma_{\rm T}^{({k= 1 \sim M})}= 1/T_2^{(e)}$. }
	\label{fig:sim_7}
\end{figure}

 We show the case for $M=7$ in Fig.~\ref{fig:sim_7}. $p_{\uparrow, k}$ approaches 0.16
 regardless of $\omega_k'$ and $g_k$. It
 converges to
 non-zero values because of 
 $\gamma_{\rm L}^{(k)} \ne 0$.
 Due to a thermal relaxation process, the state of the electron spins will be a Gibbs state in a natural environment, and its excited-state population is
$p_\uparrow = 0.47$ at 1~mT and 10~mK 
(a typical operation temperature of a FQ) environment. 
This means that
our cooling
scheme with the FQ
is especially useful when we perform an ESR with the FQ:
A sensitivity of ESR measurements is proportional to $p_\downarrow - p_\uparrow$ \cite{wertz2012electron,miyanishi2020architecture},
and so
the sensitivity 
of ESR with our polarization scheme leads 10 times better than the conventional one 
without active cooling.  
Also, it is worth mentioning that the actual temperature of the electron spins 
in the dilution refrigerator might be 50~mK or more and not 10~mK 
\cite{budoyo2018electron,toida2019electron} 
because $T_1$ of the electron spins is large
\cite{T1BudoyoBottleSCI2018}. 
Moreover, 
an interval between measurements in the standard
ESR should be a few time
larger than  $T_1$ of the electron spins.
Therefore, as $T_1$ becomes longer, our approach does more efficient 
than the conventional one.

\section{Conclusion}
\label{summary}
In conclusion, we propose a scheme to polarize electron spins with a superconducting flux qubit (FQ).
Since we cannot apply large magnetic fields for the FQ to work,
there is a large energy gap between the electron spins and FQ.
To achieve a strong interaction between them, we adopt a spin-lock technique for the FQ. 
A Rabi frequency of the FQ can be as small as resonance frequencies of the electron spins
and thus the efficient energy transfer between them can occur.
 We find that homogeneous electron spins 
without any decoherence cannot be cooled down to the ground state with the FQ, 
because the electron spins in dark states cannot be coupled to the FQ. 
Interestingly, 
dephasing on the electron spins 
(usually they are not avoidable in experiments) allows them  to escape 
from the dark states. We show that the electron spins can be polarized in realistic conditions 
by using our scheme. 

\section*{Acknowledgment }

This work was supported by Leading Initiative for
Excellent Young Researchers MEXT Japan, JST
presto (JPMJPR1919) Japan, JSPS Grants-in-Aid
for Scientific Research (21K03423),  and CREST (JPMJCR1774). 

\section*{Appendix}

\subsection{Proof of Eq.~\eqref{eq:ActionStepII}}

We, first,  prove
\begin{equation}
	\label{eq:permutation_roh}
	P\rho_{j,m}P^{\dagger}=\rho_{j,m}
\end{equation}
for any permutation matrix $P$.
Note that any permutation matrix $P$ satisfies
\begin{align}
	[P,S_{z}]=0,\,\,
	[P,S^2]=0,
\end{align}
where $S^2 = S_x^2+S_y^2+S_z^2$. 
This is directly proved by the permutation invariance $PS_{z}P^{\dagger}=S_{z}$ and $PS^{2}P^{\dagger}=S^{2}$ in the following way:
\begin{align}
	PS_{z}=PS_{z}P^{\dagger}P=S_{z}P,\nonumber\\
	PS^{2}=PS^{2}P^{\dagger}P=S^{2}P,
\end{align}
where we use $P^{-1}=P^{\dagger}$.

Let us write the action of $P$ on $|j,m,i\rangle$ as 
\begin{equation}
	P|j,m,i\rangle=\sum_{j',m',i'}P^{j,m,i}_{j',m',i'}|j',m',i'\rangle.
\end{equation}
We consider the explicit form of $P^{j,m,i}_{j',m',i'}$. 
By considering the following fact,
\begin{align}
	0=&[P,S_{z}]|j,m,i\rangle=PS_{z}|j,m,i\rangle-S_{z}P|j,m,i\rangle \nonumber\\
	=&m P|j,m,i\rangle-S_{z}\sum_{j',m',i'}P^{j,m,i}_{j',m',i'}|j',m',i'\rangle \nonumber\\
	=&m P|j,m,i\rangle-\sum_{j',m',i'} m' P^{j,m,i}_{j',m',i'}|j',m',i'\rangle \nonumber\\
	=&\sum_{j',m',i'}(m-m') P^{j,m,i}_{j',m',i'}|j',m',i'\rangle.
\end{align}
This implies that $P^{j,m,i}_{j',m',i'}$ has the form of $\delta^{m}_{m'}\tilde{P}^{j,i}_{j',i'}$.
By using the commutation relation about $S^{2}$, we can prove that $P^{j,m,i}_{j',m',i'}$ has the form of 
$\delta^{m}_{m'}\delta^{j}_{j'}\tilde{\tilde{P}}^{i}_{i'}$ in the same manner as above.
Thus, $P$ is a block-diagonal matrix with respect to the basis $\{ |j,m,i\rangle \}_{j,m,i}$.
The explicit form of $P$ is given as
\begin{equation}
	P= \sum_{j,m,i,i'} \tilde{\tilde{P}}^{i}_{i'} |j,m,i\rangle \langle j,m,i'|.
\end{equation}
because $P$ is a unitary matrix, $\tilde{\tilde{P}}^{i'}_{i}$ is also unitary, i.e., 
$\sum_{i''} \tilde{\tilde{P}}^{i}_{i''} \left( \tilde{\tilde{P}}^{i'}_{i''}\right)^*=\delta_{i,i'}$.
Then we can explicitly show,
\begin{align}
	&P\rho_{j,m}P^{\dagger}\nonumber\\
	&=\sum_{j_{1,2},m_{1,2},i_{1,2,3,4}} \tilde{\tilde{P}}^{i_{1}}_{i_{2}} 
	|j_{1},m_{1},i_{1}\rangle \langle j_{1},m_{1},i_{2}| 
	\nonumber \\
	 & \quad \quad \rho_{j,m} \left( \tilde{\tilde{P}}^{i_{4}}_{i_{3}} \right)^* 
	|j_{2},m_{2},i_{3}\rangle \langle j_{2},m_{2},i_{4}|\nonumber\\
	&=\sum_{i,i_{1},i_{4}} \tilde{\tilde{P}}^{i_{1}}_{i} \left( \tilde{\tilde{P}}^{i_{4}}_{i} \right)^*
	|j,m,i_{1}\rangle \langle j,m,i_{4}|\nonumber\\
	&=\sum_{i} |j,m,i\rangle \langle j,m,i|=\rho_{j,m}.
\end{align}
We now proved Eq.~\eqref{eq:permutation_roh}, 
which means that all the diagonal elements of $\rho_{j,m}$ represented 
in the binary basis ($|\underbrace{11\cdots1}_{n/2+m}\underbrace{00\cdots0}_{n/2-m}\rangle$ and all its permutations) is identical, that is,
\begin{equation}
	\rho_{j,m}=\frac{d_{j}}{{}_{M}C_{m+\frac{M}{2}}} |0\rangle \langle 0| \otimes 
	\begin{pmatrix}
		1&. &.&\ldots&. \\
		.&1&.&\ldots&. \\
		.&.&1&\ldots&. \\
		\vdots&\vdots&\vdots&\ddots&\vdots\\
		.&.&.&\ldots&1 \\
	\end{pmatrix}
	.
	\label{eq:identical}
\end{equation} 
This matrix is a ${}_{M}C_{m+\frac{M}{2}} \times {}_{M}C_{m+\frac{M}{2}}$ matrix while its matrix rank is $d_{j}$.
The effect of independent dephasing ${\cal E}_{\rm II}$ makes the non-diagonal elements of this matrix be $0$.
Thus, after Step II, the density matrix $\rho_{j,m}$ becomes ${\cal E}_{\rm II}(\rho_{j,m})$ given as
\begin{equation}
	{\cal E}_{\rm II}(\rho_{j,m})=\frac{d_{j}}{{}_{M}C_{m+\frac{M}{2}}} |0\rangle \langle 0| \otimes 
	\begin{pmatrix}
		1&0 &0&\ldots&0 \\
		0&1&0&\ldots&0 \\
		0&0&1&\ldots&0 \\
		\vdots&\vdots&\vdots&\ddots&\vdots\\
		0&0&0&\ldots&1 \\
	\end{pmatrix}
	,
	\label{eq:identical2}
\end{equation}
which is the identity matrix of the space spanned by the binary basis with fixed $m$.
Although the above matrix is represented in the binary basis, because the identity matrix is invariant under any unitary transformation on this space, Eq. (\ref{eq:identical2}) can be rewritten as
\begin{equation}
	{\cal E}_{\rm II}(\rho_{j,m})=\frac{d_{j}}{_{M}C_{m+\frac{M}{2}}} |0\rangle \langle 0| \otimes 
	\sum^{M/2}_{s=|m|}\sum^{d_{s}}_{i=1}|s,m,i\rangle\langle s,m,i|.
\end{equation}
Thus, Eq. (\ref{eq:ActionStepII}) is proved.
\\

\bibliographystyle{apsrev4-2}
\bibliography{7mylibrary}
\end{document}